\documentclass[11pt,cmr, a4paper]{article}
\usepackage{moriond,epsfig}

\def\be{\begin{equation}}
\def\ee{\end{equation}}
\def\bea{\begin{eqnarray}}
\def\eea{\end{eqnarray}}

\def\Blos{B_{\rm los}}
\def\mulos{\mu_{\rm los}}
\def\cs{c_{\rm s}}
\def\cso{c_{\rm s0}}
\def\va{V_{\rm A}}
\def\Alf{Alfv\'en~}
\def\sigv{\sigma_v}
\def\signi{\Sigma_{\rm n,0}}
\def\sign{\Sigma_{\rm n}}
\def\mui{\mu_0}

\def\lammax{\lambda_{\rm T,m}}
\def\la{\mathrel{\hbox{\rlap{\hbox{\lower4pt\hbox{$\sim$}}}\hbox{$<$}}}}
\def\ga{\mathrel{\hbox{\rlap{\hbox{\lower4pt\hbox{$\sim$}}}\hbox{$>$}}}}

\begin{document}
\vspace*{4cm}
\title{MAGNETIC FIELDS IN MOLECULAR CLOUD CORES}

\author{SHANTANU BASU}
\address{Department of Physics and Astronomy, 
University of Western Ontario,\\
London, Ontario N6A 3K7, Canada}
\maketitle\abstracts{ 
Observations of magnetic field strengths imply that molecular cloud
fragments are individually close to being in a magnetically critical
state, even though both magnetic field and column density measurements
range over two orders of magnitude. The turbulent pressure also 
approximately balances the self-gravitational pressure. These
results together mean that the one-dimensional velocity dispersion
$\sigv$ is proportional to the mean \Alf speed of a cloud 
$\va$. Global models of MHD turbulence in a molecular cloud show that
this correlation is naturally satisfied for a range of different
driving strengths of the turbulence. For example, an increase of
turbulent driving causes a cloud expansion which also increases
$\va$. Clouds are in a time averaged balance but exhibit large
oscillatory motions, particularly in their outer rarefied regions.
We also discuss models of gravitational fragmentation in a sheet-like
region in which turbulence has already dissipated, including
the effects of magnetic fields and ion-neutral friction.
Clouds with near-critical mass-to-flux
ratios lead to subsonic infall within cores, consistent with some
recent observations of motions in starless cores. Conversely, significantly
supercritical clouds are expected to produce extended supersonic infall.
}
\noindent
{\small¥{\it Keywords}: ISM: clouds - ISM: magnetic fields - MHD - stars: formation - turbulence }

\section{Magnetic field data}
\label{data}

When discussing magnetic fields in molecular clouds, a useful
starting point is to look at the confirmed detections of magnetic
field strength using the Zeeman effect. Other methods using measurements
of polarized emission and the 
Chandrasekhar-Fermi method are just beginning to be applied to molecular
clouds, but are characterized by larger uncertainties.
We look at data encompassing 15 confirmed detections compiled
by Crutcher (1999), one more detection (in L1544) by Crutcher \& Troland 
(2000), and one (in RCW38) by Bourke et al. (2001). 
The last one does not have a density estimate, so is used only in 
Fig. 1a below.
It emerges that the data satisfy two
independent correlations, as first noted by Myers \& Goodman (1988)
from data available at that time.
We cast these correlations in the following manner. 

\begin{figure}
\resizebox{\hsize}{!}{\includegraphics{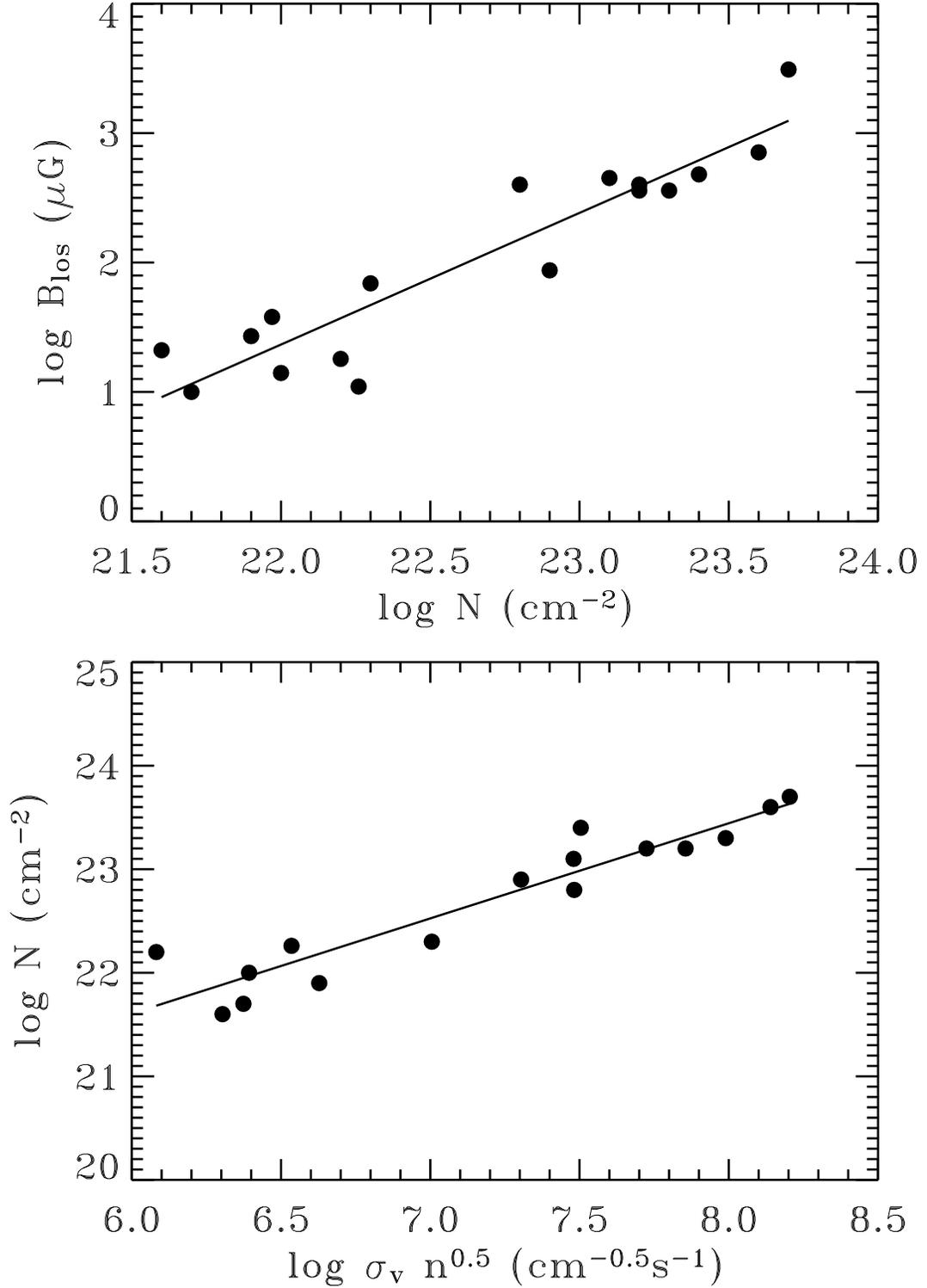}}
\smallskip
\caption{{\it Top}: Plot of $\log \Blos$ versus $\log N$ for the sample of
17 clouds with confirmed magnetic field detections (see text). The solid line
is a least squares best fit to the data.
{\it Bottom}: Plot of $\log N$ versus $\log \sigv n^{0.5}$ for the 
same sample of clouds (minus RCW38). 
The solid line is a least squares best fit to the data.
}
\end{figure}

Figure 1a shows that there is a clear correlation between $\Blos$, 
the line-of-sight
component of the magnetic field that is measurable by the Zeeman 
effect, and the column density $N$ (by definition also a line-of-sight 
value), in over two orders of magnitude variation in each quantity. 
This is expected from the relation
\be
\frac{\Sigma}{B} \equiv \mu \, (2 \pi G^{1/2})^{-1}
\ee
(where $\Sigma = m N$, in which $m$ is the mean molecular mass)
if the dimensionless mass-to-flux ratio $\mu$, measured in units of the 
critical value $(2 \pi G^{1/2})^{-1}$ (Nakano \& Nakamura 1978), 
is approximately constant from cloud to cloud.
The solid line is the least squares best-fit to $\log \Blos$ versus 
$\log N$. It has an estimated slope $1.02 \pm 0.10$, consistent with
the expected slope of unity. The average measured mass-to-flux ratio is
$\langle \mulos \rangle = \langle \Sigma/\Blos \rangle \times 2 \pi G^{1/2} 
= 3.25$. Since the measured field $\Blos$ is related to the full magnetic
field strength $B$ by $\Blos = B \cos \theta$, then if the magnetic fields
have a random set of inclinations $\theta$ to the line-of-sight, 
we would expect that $\langle B \rangle = 2 \langle \Blos \rangle$. 
Assuming that $N$ is the same for all lines of
sight, we then expect the average total mass-to-flux ratio to be
$\langle \mu \rangle = \langle \Sigma/B \rangle \times 2 \pi G^{1/2} = 1/2 \langle \mulos \rangle = 1.63$.
We also note that if the clouds are preferentially flattened along
the magnetic field direction, the column density parallel to the 
magnetic field $N_{\parallel} = N \cos \theta$.
Since $N_{\parallel}$ is the relevant column density for calculating the
mass-to-flux ratio, in this case we get $\langle \mu \rangle = 
\langle \Sigma_{\parallel}/B \rangle \times 2 \pi G^{1/2} = 1/3 \langle \mulos \rangle = 1.08$, using an angle-averaging process (see Crutcher 1999).
All in all, it is a remarkable feature that molecular clouds are
so close to a magnetically {\it critical} state over a wide range
of observed length scales and densities. 

A second correlation is between the self-gravitational pressure at
the midplane of a cloud and the internal turbulent pressure.
One may imagine that a cloud settles into such a state by establishing
approximate force balance along magnetic field lines.
In this case, we expect
\be
\rho_0 \, \sigv^2 = \frac{\pi}{2} G \, \Sigma^2,
\ee
where $\rho_0$ is the density at the midplane, and $\sigv$ is the total
(thermal and non-thermal) one-dimensional velocity dispersion.
We have assumed that the effect of confining external pressure is small
compared to the self-gravitational pressure. Since the mean density
$\rho$ may be related to $\rho_0$ by some multiplicative constant,
we expect that 
\be
N \propto \sigv n^{1/2},
\label{linesize}
\ee
where $n = \rho/m$ is the mean number density.
Figure 1b shows that this relation is indeed valid  
for our cloud sample. The least squares best-fit
yields a slope $0.92 \pm 0.09$, again consistent with unity.
We note in passing that equation (\ref{linesize}) is the generalized 
form of the well-known linewidth-size relation for molecular clouds;
if $N \propto n R \approx {\rm constant}$ (unlike this sample) 
for a sample of clouds
of different radii $R$, then $\sigv \propto n^{-1/2} \propto R^{1/2}$.

Since $B \propto N$ and $N \propto \sigv n^{1/2}$,
it is clear that we expect 
\be
B \propto \sigv n^{1/2}. 
\ee
Figure 2 shows this correlation from the data using $\Blos$ instead of $B$.
This relation is equivalent to 
$\sigv \propto \va$, where $\va \equiv B/\sqrt{4 \pi \rho}$ is the 
mean \Alf speed of the cloud, calculated using the mean density
$\rho$. 
Our best fit (solid line) yields a slope $1.03 \pm 0.09$.
The average ratio $\langle \sigv/\va \rangle  = 0.54$, if we again use
$\langle B \rangle  = 2 \langle \Blos \rangle$.

\begin{figure}
\resizebox{\hsize}{!}{\includegraphics{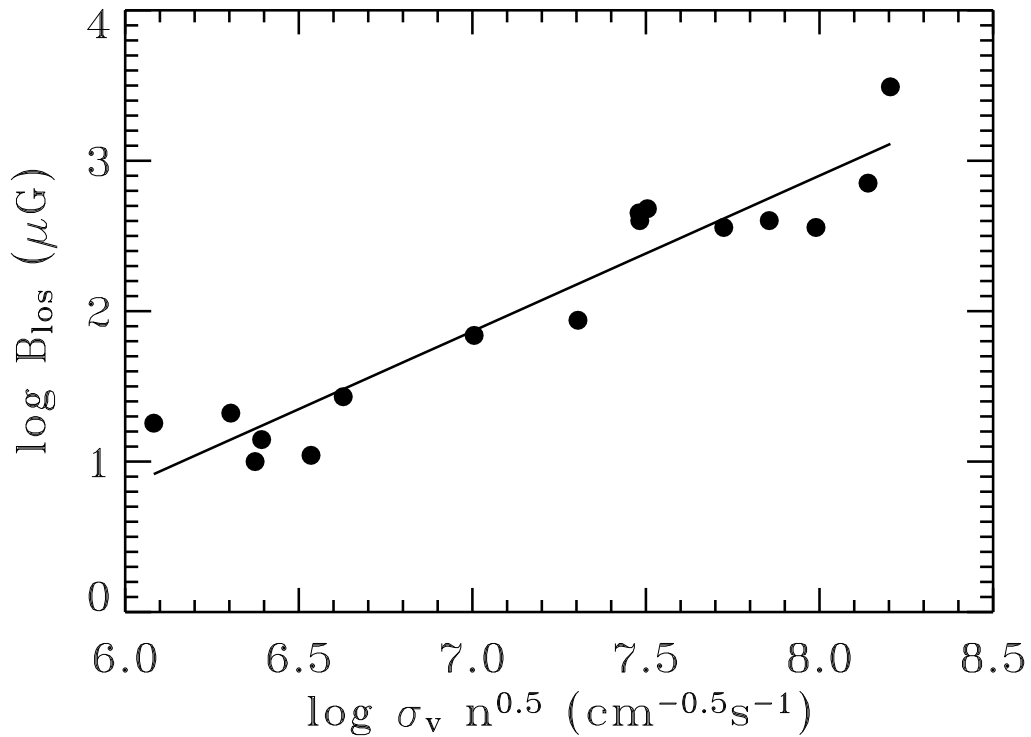}}
\smallskip
\caption{
Plot of $\log \Blos$ versus $\log \sigv n^{1/2}$ for the
same clouds as in Fig. 1b. The solid line is the least squares best
fit.
}
\end{figure}

It is also interesting to note here that derived relation
$\sigv \propto \va$ does {\it not} necessarily imply that the turbulence
consists of Alfv\'enic motions. We have only assumed a near critical
mass-to-flux ratio and any unspecified turbulent motions.
The relationship is a reflection of the global properties of a cloud,
and follows from the virial relations (Myers \& Goodman 1988).
Indeed, \Alf waves alone might lead to material motions
that are significantly sub-Alfv\'enic. For example, linear
\Alf waves obey the relation $\delta v = \va \delta B/B$, where 
$\delta v$ and $\delta B$ are the amplitudes of fluctuations in the
material speed and magnetic field, respectively. Thus we see that
$\delta v \ll \va$ if $\delta B/B \ll 1$.

\section{A Model for MHD Turbulence}

Kudoh \& Basu (2003) have presented a numerical model of MHD turbulence 
in a stratified, bounded, one-dimensional cloud. The model is 1.5 
dimensional, meaning that vector quantities have both $y$ and $z$ 
components, but can only vary in the $z$-direction.
It is a {\it global} model of turbulence, in contrast to a {\it local}
periodic box numerical model. In this model, the cloud stratification
can be modeled, and the mean cloud density $\rho$ can change with time.

\begin{figure}
\centerline{\includegraphics[width=84mm]{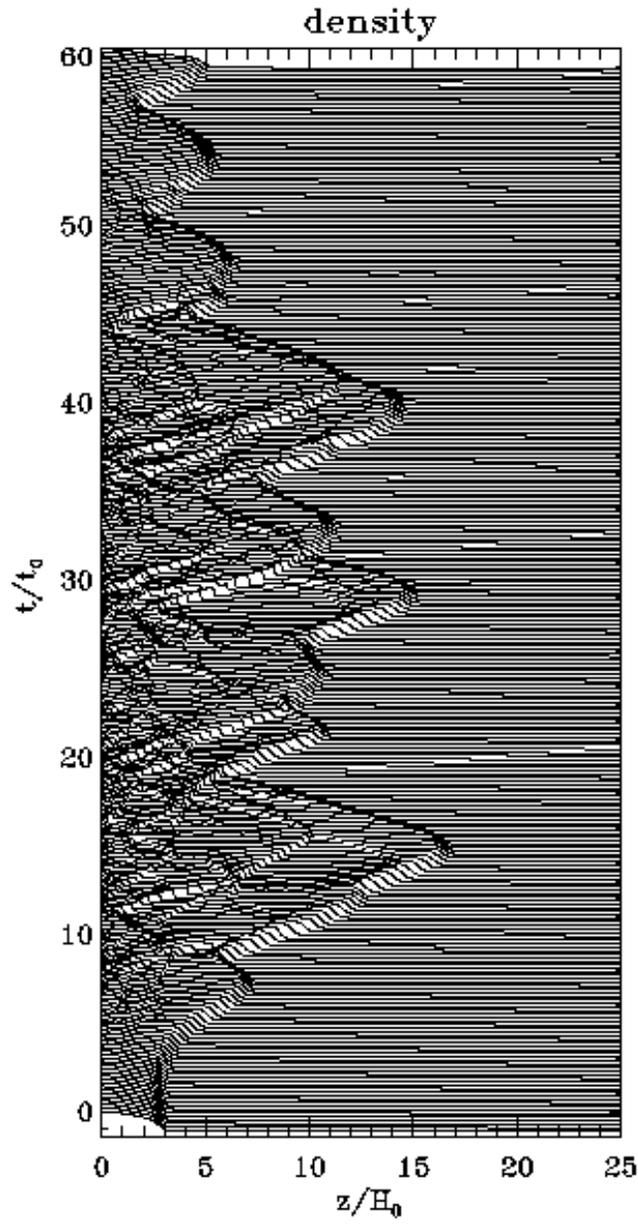}}
\smallskip
\caption{Time evolution of the density in a global model
of MHD turbulence (Kudoh \& Basu 2003). 
The density versus $z/H_0$ at various 
times are stacked with time, with time increasing upwards
in uniform increments of $0.2t_0$.
}
\end{figure}

The model initial condition is a hydrostatic
equilibrium between thermal pressure and self-gravity in a cloud that
is bounded by an external high temperature medium.
The initial state of the cloud is a truncated
(at about $z=3H_0$) Spitzer equilibrium density profile
$\rho(z)=\rho_0 \, {\rm sech}^2 (z/H_0)$, in which
$H_0=\cso / \sqrt{2\pi G\rho_0}$ and $\cso$ is the isothermal
sound speed of the cloud.
Isothermality is maintained for each Lagrangian fluid element.
A sinusoidal driving force of dimensionless amplitude
$\tilde{a}_d$ (see Kudoh \& Basu 2003 for details) is introduced near 
the midplane of the cloud
and the dynamical evolution of the vertical structure
of the cloud is followed. The cloud is characterized by a 
critical mass-to-flux ratio ($\mu = 1$).
Figure 3 shows the time evolution of the density presented by 
Kudoh \& Basu (2003).
The density plots at various times are stacked with time
increasing upward in uniform increments of $0.2t_0$,
where $t_0=H_0/c_{s0}$.
Because the driving force increases linearly with time up to
$t=10t_0$, the density changes gradually at first.
After $t=10t_0$,
the density structure shows many shock waves
propagating in the cloud, and significant upward and downward
motions of the outer portion of the cloud, including the
temperature transition region.
After terminating the driving force at $t=40t_0$,
the shock waves are dissipated in the cloud and
the transition region moves back toward the initial position,
although it is still oscillating.
A stronger driving force (larger $\tilde{a}_d$) causes a larger
turbulent velocity, which results in a more dynamic evolution of
the molecular cloud, including stronger shock waves, and larger
excursions of the cloud boundary.

Figure 4a shows the time averaged velocity dispersions
$\langle \sigma^2 \rangle_t^{1/2}$
of different Lagrangian fluid elements for different
strengths of the driving force, as a function of the time averaged height
$\langle z \rangle_t$. The open circles correspond to
Lagrangian fluid elements close to the cloud edge, while the
filled circles represent fluid elements near the half-mass position
of the cloud. 
Each circle corresponds to a different value of $\tilde{a}_d$,
with increasing $\tilde{a}_d$ generally resulting in increasing 
$\langle z \rangle_t$.
The dotted line shows
\begin{equation}
\langle \sigma^2 \rangle_t^{1/2} \propto \langle z \rangle_t^{0.5},
\end{equation}
and reveals that the model clouds are in 
a time-averaged equilibrium state. The relation is also
consistent with the well-known observational
linewidth-size relation
of molecular clouds (e.g., Larson 1981; Solomon et al. 1987).

\begin{figure}
\resizebox{\hsize}{!}{\includegraphics{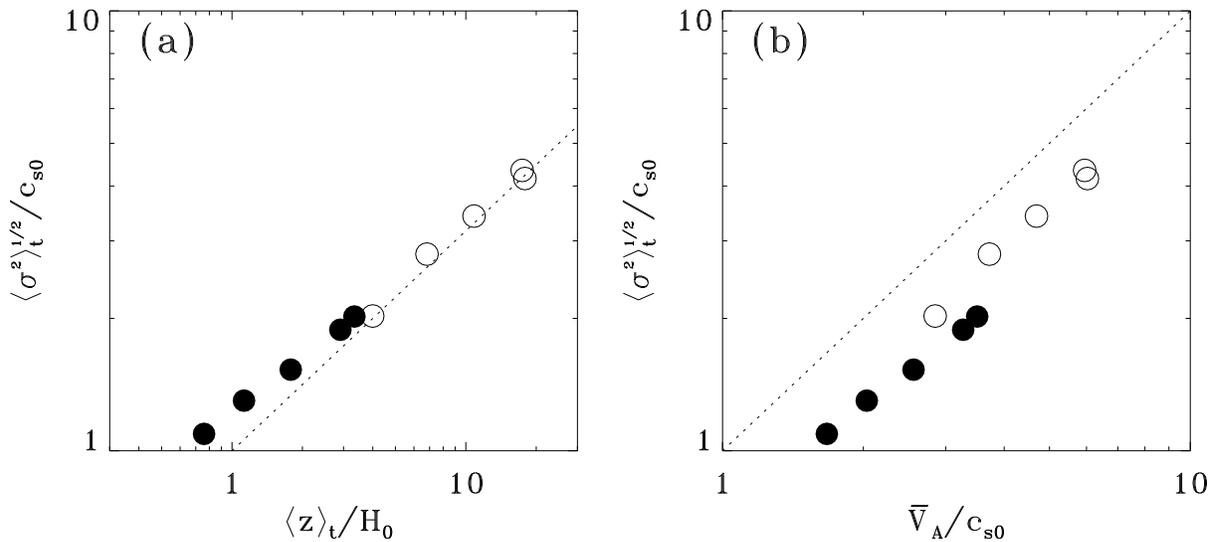}}
\smallskip
\caption{
Global properties of an ensemble of clouds with different turbulent driving
strengths $\tilde{a}_d$.
(a) Time averaged velocity dispersions $\langle \sigma^2 \rangle_t^{1/2}$ of
different Lagrangian fluid elements for different $\tilde{a}_d$,
as a function of time averaged positions $\langle z \rangle_t^{0.5}$.
The open circles correspond to Lagrangian fluid elements
whose initial positions are $z/H_0=2.51$, close to the cloud edge.
The filled circles correspond to Lagrangian fluid elements
whose initial positions are $z/H_0=0.61$,
approximately the cloud half-mass position.
The dotted line is the relation 
$\langle \sigma^2 \rangle_t^{1/2} = \langle z \rangle_t^{0.5}$.
(b) $\langle \sigma^2 \rangle_t^{1/2}$ versus mean \Alf speed
$\va \equiv B/\sqrt{4 \pi \rho}$, where $\rho$ is the average density.
The dotted line shows
$\langle \sigma^2 \rangle_t^{1/2} = \va$.
All quantities are normalized to the isothermal sound speed $\cso$ in the
cloud.
}
\end{figure}

Figure 4b plots $\langle \sigma^2 \rangle_t^{1/2}$
versus the mean Alfv\'en speed $\va$ 
for individual Lagrangian fluid elements.
The dotted line shows
\begin{equation}
\langle \sigma^2 \rangle_t^{1/2} \propto \va,
\end{equation}
and reveals that the simulations result in a good correlation between
$\langle \sigma^2 \rangle_t^{1/2}$ and 
$\va \equiv B_0/{\sqrt{4\pi\rho}}$, where
$\rho=\Sigma/(2 \langle z \rangle_t)$
is the mean density and $\Sigma$ is the column density for each
Lagrangian element having mean position $\langle z \rangle_t$.
This relation is essentially the same as the observational correlation
$B \propto \sigv n^{1/2}$ presented in \S\ 1.

It is worth noting here that the motions inside the cloud are 
overall slightly sub-Alfv\'enic, and highly sub-Alfv\'enic
in the rarefied envelopes of the stratified clouds, where the local
\Alf speed can be very high.
Furthermore, very strong driving {\it fails} to produce 
super-Alfv\'enic motions, 
due to the ability of the cloud to expand and lower its 
density, thus increasing $\va$. A natural time-averaged balance is always 
established in which
$\sigv \approx 0.5 \va$; both $\sigv$ and $\va$ are variable quantities,
unlike in a periodic box simulation where they may be held fixed.

\section{Fragmentation of a Magnetized Cloud}

Basu \& Ciolek (2004) have modeled the evolution of a two-dimensional 
region perpendicular to the mean magnetic field direction, using the 
thin-disk approximation. This is a non-ideal MHD simulation which includes
the effect of ambipolar diffusion (ion-neutral drift), in a region
that is partially ionized by cosmic rays.
Physically, this model is complementary to that of Kudoh \& Basu (2003)
in that it models the other two dimensions (perpendicular to the mean
magnetic field), in a sub-region of a cloud where turbulence has 
largely dissipated. Ion-neutral friction is also expected to be more
efficient in subregions of clouds where the background ultraviolet 
starlight cannot
penetrate (McKee 1989), and the ionization fraction is therefore much
lower. Indeed, MHD turbulent motions may also be preferentially 
damped in regions with lower ionization fraction (e.g., Myers \&
Lazarian 1998).
The two-dimensional computational domain is modeled with periodic 
boundary conditions and has an initially uniform column density 
$\sign$ (``n''denotes neutrals) and vertical magnetic field $B_z$. 
Small white-noise perturbations
are added to both quantities in order to initiate evolution.

Figure 5 shows the contours of $\sign/\signi$ and the velocity vectors
for a model with critical initial mass-to-flux ratio ($\mu_0 = 1$), at
a time when the maximum value of $\sign/\signi \approx 10$. The time
is $t = 133.9 \, t_0$, where $t_0 = \cs / (2 \pi G \signi) = 
2.38 \times 10^5$ yr for an 
initial volume density $n_{\rm n,0} = 3 \times 10^{3}$ cm$^{-3}$.
Star formation is expected to occur very shortly afterward in the 
peaks due to the very short dynamical times in those regions, which are
now magnetically supercritical. Although it takes a significant time 
$\approx 3 \times 10^7$ yr for the peaks to evolve into the runaway phase, 
it is
worth noting that nonlinear perturbations would result in lesser times.
The contours of mass-to-flux ratio $\mu(x,y) = \sign(x,y)/B_z(x,y)
\times 2 \pi G^{1/2}$ (now nonuniform due to ion-neutral drift) also
reveal that regions with $\sign/\signi > 1$ are typically supercritical while 
regions with $\sign/\signi < 1$ are typically subcritical 
(Basu \& Ciolek 2004).
This means that ambipolar diffusion leads to flux redistribution that 
naturally creates {\it both} supercritical
and subcritical regions in a cloud that is critical ($\mu_0 = 1$) 
overall. A distinguishing characteristic of the critical model is that 
the infall motions are {\it subsonic}, both inside the core and outside,
with maximum values $\approx 0.5 \cs \approx 0.1$ km s$^{-1}$ found 
within the cores. This is consistent with detected infall motions in
some starless cores, specifically in L1544 (Tafalla et al. 1998;
Williams et al. 1999).
The core shapes are mildly non-circular in the plane, and triaxial when 
height $Z$ consistent with vertical force balance is 
calculated. However, the triaxial shapes are closer to oblate than
prolate since the $x$- and $y$- extents are roughly comparable and both much
greater than the extent in the $z$-direction.

Figure 6 shows the contours of $\sign/\signi$ and the velocity vectors
for a model with $\mu_0 = 2$, i.e., significantly supercritical
initially. The distinguishing characteristics of this model are the
relatively short time ($t = 17.6 \, t_0 \approx 4 \times 10^6$ yr) 
required to reach a maximum $\sign/\signi \approx 10$, the supersonic 
infall motions within
cores, and the more elongated triaxial shapes of the cores.
Note that the velocity vectors in Fig. 6 have the same normalization
as in Fig. 5. The extended supersonic infall (on scales $\la 0.1$ pc)
provides an observationally distinguishable difference between
clouds being critical or significantly supercritical.

\begin{figure}
\centerline{\includegraphics[width=84mm]{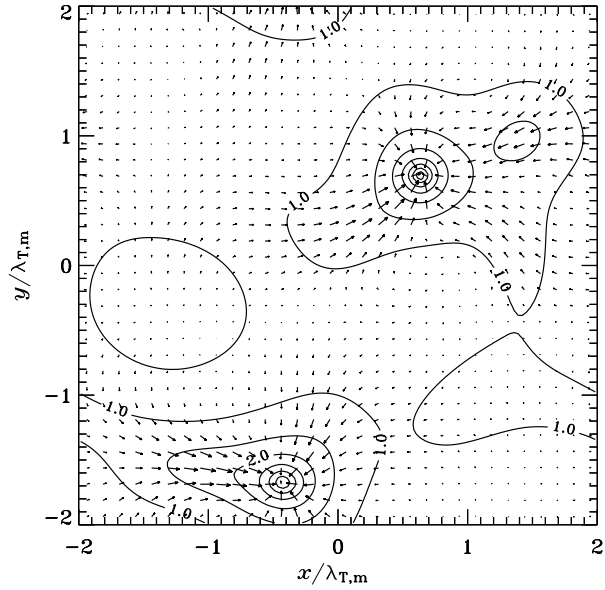}}
\smallskip
\caption{Fragmentation in the critical model ($\mui = 1$) of 
Basu \& Ciolek (2004). The data are shown when the maximum
column density $\approx 10 \, \signi$. Lines represent contours of 
normalized column density $\sign(x,y)/\signi$, 
spaced in multiplicative increments of $2^{1/2}$, 
i.e., [0.7,1.0,1.4,2,2.8,...]. Also
shown are velocity vectors of the neutrals; the distance between tips of
vectors corresponds to a speed $0.5 \, \cs$. 
The positions $x$ and $y$ are normalized to $\lammax$, the wavelength of
maximum growth rate for linear perturbations in a nonmagnetic sheet.
}
\end{figure}

\begin{figure}
\centerline{\includegraphics[width=84mm]{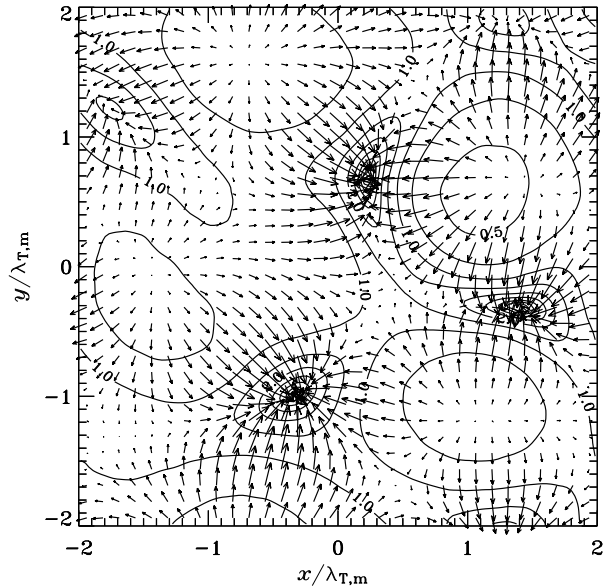}}
\smallskip
\caption{Fragmentation in a supercritical model ($\mui = 2$) of 
Basu \& Ciolek (2004). 
The data are shown when the
maximum column density $\approx 10 \, \signi$.
Lines represent contours of normalized column density $\sign(x,y)/\signi$,
spaced in multiplicative increments of $2^{1/2}$. Also
shown are velocity vectors of the neutrals; they have the same normalization
as in Fig. 5. Note the significantly more rapid motions in this case.
}
\end{figure}

\section{Discussion and Conclusions}

We have seen that any cloud that has a balance of self-gravitational
pressure and turbulent pressure, and also has a large-scale magnetic
field such that $\mu \sim 1$ will satisfy the relation
$B \propto \sigv n^{1/2}$ ($\sigma \propto \va$). 
Observed cloud fragments satisfy this correlation very well.
Numerical experiments (Kudoh \& Basu 2003) modeling the global effects
of internal MHD turbulence show that clouds evolve in an oscillatory
fashion (with the outer parts making the largest excursions) 
but satisfy the above correlation in a time-averaged sense.
The temporal averaging in that model may also be akin to a spatial 
averaging through many layers of cloud material along the line of sight.
We emphasize that the observations and numerical
simulations imply that
clouds can readjust to {\it any} level of internal turbulence in 
such a way that $\sigv$ and $\va$ come into approximate balance
(specifically $\sigv \approx 0.5 \va$).
Unlike the sound speed $\cs$ in an isothermal cloud, the mean
\Alf speed $\va$ in a self-gravitating cloud is not a fixed quantity, 
and varies in space and time, as the cloud expands and contracts.

Molecular cloud fragments seem to represent an ensemble of objects
with varying levels of turbulent support, but which have
a near-critical mass-to-flux ratio. The data has sometimes been
suggested to be consistent with the relation $B \propto n^{1/2}$.
Such a relation is expected for the contraction of a cloud that is
flattened along the magnetic field direction, if flux-freezing holds.
It is roughly satisfied by the data (see Crutcher 1999) given that $\sigv$ 
only varies by 
one order of magnitude while $B$ and $n$ have much larger
variations. However, as shown by Basu (2000), the correlation is
much better for $B \propto \sigv n^{1/2}$. We believe that the proper
interpretation is that the cloud fragments all have near-critical
mass-to-flux ratio and varying levels of internal turbulence.
These clouds do not represent a direct evolutionary sequence since the
sizes and masses of the objects differ by many orders of magnitude,
e.g., the sizes range from 22.0 pc down to 0.02 pc, and masses from
$\sim 10^6$ M$_{\odot}$ down to $\sim 1$  M$_{\odot}$!

In regions where turbulence has largely dissipated, one may expect 
a gravitational fragmentation process regulated by (non-ideal) MHD
effects, as modeled by Basu \& Ciolek (2004). We note that the
non-turbulent models also satisfy $\sigv \sim \va$, where 
$\sigv = \cs$ in this case, since it is essentially thermal pressure which 
balances the gravitational pressure along field lines.
We have found that the fragmentation process of a significantly supercritical
cloud may be ruled out in the context of current star formation
in e.g., the Taurus molecular cloud, due to the lack of observed
supersonic infall (Tafalla et al. 1998; Williams et al. 1999).
Velocity fields provide an interesting distinguishing 
characteristic of various levels of magnetic support.
Cloud core shapes are invariably triaxial, and closer to oblate rather
than prolate. The observed distribution of cloud core shapes, which 
imply triaxial but more nearly oblate objects (Jones, Basu, \& Dubinski
2001) can be naturally understood using these kind of models, although
more complete models will need to be truly three-dimensional and
include internal turbulent support.
The critical model of Basu \& Ciolek (2004) also shows that
flux and mass redistribution naturally creates both supercritical regions and
subcritical envelopes. Mass redistribution in flux tubes is a key
feature of gravitationally driven ambipolar diffusion, as emphasized
long ago by Mouschovias (1978). Detailed 
targeted observations of the inter-core medium are necessary in order to
identify the putative subcritical envelopes.
All in all, the outcome of gravitational fragmentation in a non-ideal MHD
environment in which $\mu \approx 1$ may hold many surprises. We
are just beginning to explore the rich physics of such systems.

Looking forward, we must grapple with several key questions about the
role of the magnetic field and star formation in general.
Are the triaxial shapes of cores an important factor in binary or
multiple system formation? This will require high-resolution 
MHD simulations of nonaxisymmetric cores. 
Do the different rates of infall in 
subcritical, critical, and supercritical clouds actually affect the
final outcome? We have heard at this meeting that star formation in
many environments (e.g., starbursts) can be quite efficient. Perhaps
supercritical fragmentation was important in the past history of the
Galaxy and in external galaxies, while the relatively inefficient
current day star formation in the Galaxy is the result of critical
or subcritical fragmentation. We need to quantify to what extent
a subcritical or critical cloud can limit star formation through
subcritical envelopes which have an inability or lack of
available time to form stars.
Simulations which go much further ahead in time, and include the 
feedback effect of the first generation of stars, can answer these questions.
At this point, we are not sure to what extent stellar masses are determined by
(1) a finite mass reservoir due to envelopes supported by 
magnetic and/or turbulent support, and/or (2) feedback from outflows.
Future observations and numerical models should resolve this issue.

\section*{Acknowledgments}
I thank Glenn Ciolek, Takahiro Kudoh, Eduard Vorobyov, and James Wurster for 
their collaborative work and many stimulating discussions.
I am also very appreciative of the organizers
for a most thought-provoking and enjoyable meeting.
This work was supported by the Natural Sciences and Engineering
Research Council (NSERC) of Canada.

\end{document}